\documentclass{ws-p8-50x6-00}

\def\beq{\begin{equation}}
\def\eeq{\end{equation}}
\def\bea{\begin{eqnarray}}  \def\eea{\end{eqnarray}}
\newcommand{\noi}{\noindent}

\begin{document}

\title{$J/\psi$ suppression at SPS and RHIC energies}

\author{Elena G. Ferreiro}

\address{Departamento de F\'{\i}sica de Part\'{\i}culas, 
%Universidad de
%Santiago de Compostela, 
15706 Santiago de Compostela, Spain \\
E-mail: elena@fpaxp1.usc.es}


\maketitle

\abstracts{
The strong suppression of
the $J/\psi$
%shown by the experimental data on central Pb--Pb collisions at SPS
%energies,
is studied in the framework of hadronic and quark
gluon plasma models.
% and models that include quark
%gluon plasma formation. 
Predictions for RHIC energies are presented.}
%We also discuss the fact that, together with the possibility of
%deconfinement which produces a suppression of the $J/\psi$, there is also
%an increase of heavy flavoured pair production that should be taken into
%account a very high energies.}

\section{Introduction}
%An interesting 
%The result of the
%The last run (1998 data) of 
The NA50 collaboration data \cite{1r} on PbPb central collisions at SPS
energies
%at CERN \cite{1r}
%on the
%transverse energy ($E_T$) dependence of $J/\psi$ suppression in $PbPb$
%collisions
%,
%is the
%observation \cite{1r} 
%that the slope of the
has shown a strong suppression of 
ratio $R(E_T)$ of $J/\psi$ over Drell-Yan (DY) 
%cross-sections 
that
increases with
increasing $E_T$. This suppression 
%is stronger than the expected one due to the
%usual nuclear absorption, and 
has been proposed 
as an evidence 
%\cite{ref7} 
of the obtention of the
quark gluon plasma (QGP).
%
%In this work 
%we analize the present data 
%at SPS energies 
We present our results
in the framework of an
hadronic model \cite{capella} (without QGP)
%-without 
%considering 
%QGP formation-
%and also in the framework of a string model
and those of a 
string 
model  
%that 
%takes into account 
which includes 
percolation
of strings 
as 
a way of 
QGP 
%formation 
\cite{nosotros}.

\section{Hadronic model}
We use the model 
%introduced in 
of 
ref. 
\cite{capella}. Here, 
as
in most non-QGP models, the $J/\psi$ suppression is due to two mechanisms~:
absorption of the
pre-resonant $c\bar{c}$ pair with nucleons (the 
%so-called 
nuclear absorption)
and the
interaction of the $J/\psi$ with comovers. The corresponding $J/\psi$ survival
probabilities
are given by 
%\cite{2r}.
%
\beq
S^{abs}(b, s) = {\left \{ 1 - \exp [- A \ T_A(s) \ \sigma_{abs}] \right \}
\left \{ 1 - \exp
\left [- B\ T_B(b-s) \ \sigma_{abs}\right ]\right \} \over \sigma_{abs}^2 \ AB
\ T_A(s) \
T_B(b-s)} \ , \label{1e} \eeq
\beq S^{co}(b, s) = \exp \left \{ - \sigma_{co} \ N_y^{co}(b,
s)\   \ln \left ( {N_y^{co}(b, s) \over N_f} \right )\   \right \} \ .
 \label{2e} \eeq
\noindent
The survival probability $S^{co}$ depends on the density of comovers
$N_y^{co}(b,s)$, that we have computed using the dual parton model (DPM), 
%in the
%rapidity region of the dimuon trigger $2.9 < y_{lab} < 3.9$
and $N_f = 1.15$~fm$^{-2}$ 
%\cite{2r,4r} 
is the corresponding
density in $pp$ collisions.
%In order to compute $N_y^{co}$, a formula based on the dual
%parton model (DPM) has been used. 
%In this model the density of comovers is
%proportional to the transverse energy $E_T$. But this is only true up to the
%knee
%of the $E_T$ distribution ($E_T \sim$ 100 GeV). Beyond this value, we enter
%into the tail of
%the $E_T$ distribution - where the increase in $E_T$ is due to fluctuations.
%We need to introduce these fluctuations.

%Let us start by recalling the formulae needed to calculate the $J/\psi$
%suppression: 
%\par
%\noi 
At fixed impact parameter $b$, the $J/\psi$ cross-section is given by 
%\cite{2r}
 \beq
\label{3e}
\sigma_{AB}^{\psi}(b) = {\sigma_{pp}^{\psi} \over
\sigma_{pp}} \int d^2s \  m(b,s) \ S^{abs}(b,s) \ S^{co}(b, s)\  , \eeq
\noi where $m(b, s) = AB \  \sigma_{pp} \ T_A(s) \ T_B(b - s)$. The
corresponding one for DY pair
production is obtained from (\ref{3e}) putting $\sigma_{abs} = \sigma_{co} = 0$
(i.e. $S^{abs}
= S^{co} = 1)$ and  is proportional to $AB$. In this way we can compute the
ratio of $J/\psi$ over DY as a function of the impact parameter.
Experimentally, however, the ratio $R(E_T)$ is given
as a function of the transverse energy $E_T$.
% measured by a calorimeter, in the
%rapidity interval
%$1.1 < y_{lab} < 2.3$. 
In order to compute $R(E_T)$ we have 
%So we need 
to know the
correlation $P(E_T, b)$
between $E_T$ and impact parameter, which is given by 
%\cite{2r}
\beq
\label{4e}
P(E_T , b) = {1 \over \sqrt{2 \pi \ q\  a\ E_T^{NF}(b)}} \exp \left [ -
\left [ { E_T - E_T^{NF}(b)} \right ]^2
\over 2q \ a\ E_T^{NF}(b) \right ]\ ,
\eeq
where $E_T^{NF}(b) = q\ N_{cal}^{co}(b)$.
% -plus a term due to the
%intranuclear cascade, only important at low $E_T$.
%
%The
%$J/\psi$ and DY cross-section at fixed $E_T$ are then given by
%\beq
%\label{6e}
%{d\sigma^{\psi (DY)} \over dE_T} = \int d^2b \ \sigma_{AB}^{\psi (DY)} \ P(E_T,
%b)\ .
%\eeq
The quantity $E_T^{NF}(b)$ in eq. (\ref{4e}) does not contain fluctuations.
% -
%hence the index
%$NF$. 
In effect, if we calculate the quantity $F(E_T) = E_T/E_T^{NF}(E_T)$,
%\beq
%\label{7e}
%F(E_T) = E_T/E_T^{NF}(E_T) \ ,
%\eeq
%\noi where
%\beq \label{8e}
%E_T^{NF} (E_T) = {\int d^2b \ E_T^{NF}(b) \ P(E_T,b) \over \int d^2b \ P(E_T,
%b)}\ , \eeq
%\noi 
we see that $E_T^{NF}$ coincides with $E_T$ only up to the knee of the
$E_T$ distribution.
Beyond it, $E_T^{NF}$ is smaller than the true value of $E_T$. 
%This difference
%is precisely due
%to the $E_T$ fluctuation. 

So in order to compute the ratio $R(E_T)$ beyond the knee of the $E_T$
distribution it is necessary to introduce in $N_y^{co}$ the $E_T$ (or
multiplicity) fluctuations
responsible for the tail of the distribution.
%In order to do so, we use the experimental observation that multiplicity
%and $E_T$ distributions have approximately the same shape.
This leads to the replacement $N_y^{co}(b,s) \to N_y^{co}(b,s)\ F(E_T)$ in the
above equations.
%This leads to the following replacement 
%in eq. (\ref{2e}):
%In order to do so we multiply the density of
%comovers in eq. (\ref{2e}) by $F(E_T)$, i.e. we make the following replacement
%
%\beq \label{9e}
%N_y^{co}(b,s) \to N_y^{co}(b,s)\ F(E_T) \ . \eeq
%
In this way the results for the ratio $R(E_T)$ are unchanged below the
knee of the
distribution. 
%(see Fig.~1). 
Beyond it, the $J/\psi$ suppression is increased as
a result of the
fluctuation.
\begin{figure}[t]
\centering\leavevmode
\epsfxsize=3.75in\epsfysize=2in\epsffile{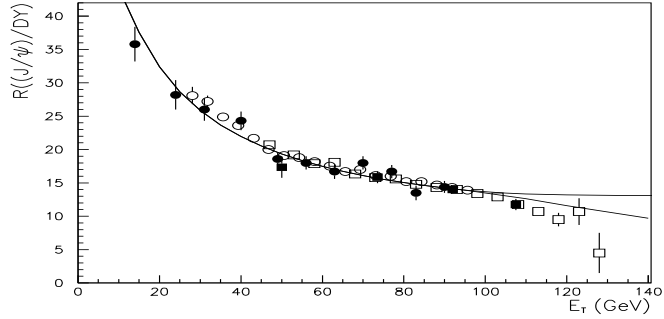}
\caption{The ratio $R(E_T)$
of $J/\psi$ over DY cross-sections, obtained with
$\sigma_{abs} = 4.5$~mb and $\sigma_{co} = 1$~mb, compared to the NA50 data.
%The different curves at low $E_T$ corresponds to
%different ways of taking into account intranuclear cascade.
The non-saturation of the ratio at $E_T > 100$ GeV is obtained when including
$E_T$ fluctuations.}
%The black points correspond to 1996 Pb-Pb data, the black squares correspond to
%1998 Pb-Pb data, the white points to 1996 analysis with minimum bias
%and the white squares to 1998 analysis with minimum bias.}
\label{figure1}
\end{figure}
\section{Percolation of strings}
In most of the hadronic models of multiparticle production, color strings
are exchanged between projectile and target. 
%These strings decay afterwards
%into the observed
%hadrons. Color 
These strings may be viewed as small areas $\pi r_0^2$, $r_0 \simeq$
0.2 fm,
in the transverse
space.
%, filled with the color field created by
%the colliding partons. 
Particles are produced via emission of $q \bar q$
pairs
in this color field.
The number of exchanged strings grows with the energy
%, the centrality 
and
the atomic number
of the colliding
particles.
%; thus it is natural to consider that
When the density of strings is high, they can overlap,
forming clusters.
%, very much like disks in continuum two-dimensional percolation
%theory. 
At a certain critical density a large cluster appears,
which signs the percolation phase transition.
% \cite{ref1,ref2,ref3,ref4}.

A cluster formed by many strings has a very high
color and therefore a very large string tension which can enhance the
$c \bar c$ pair production.
% several orders of magnitude.
% \cite{ref5,ref6}.
This effect works in the opposite direction to the Debye screening
% \cite{ref7}
which
makes that above the percolation threshold the probability of binding
the $c \bar c$ pair to form a $J/\psi$ is strongly reduced.

Here we are going to compute in a single and direct way the two oppossite
effects at
SPS, RHIC and LHC energies.

\subsection{Enhacement of $c {\bar c}$ production}
%Let us start by considering 
We consider the extension of the Schwinger formula
%\cite{ref5,ref6} 
for the production of $q- \bar q$ pairs of  mass
$m_j$ in a uniform color field with charge $g_j$, per unit space-time
volume
\begin{equation}{dN_{q- \bar q} \over dy} =  {1 \over 8 \pi^3}
\int_0^\infty d \tau \tau\int d^2x_T
|g_j E|^2\sum_{n=1}^\infty
{1 \over n^2}
\exp \Big (  -{\pi n m_j^2\over |g_j E|}\Big )\ \ .
\label{ec1}\end{equation}

The strings form clusters,
each of them with a constant color field $E_i=Q_i/S_i$,
where $Q_i$ and $S_i$ correspond to the cluster color charge and the cluster
area.
The charge and the field of each cluster before the decay, $Q_{i0}$ and
$E_{i0}$,
depend
on the number $n_i$ of strings and the area $S_1$ of each individual string
that comes into the cluster, as well as on
the total area of the cluster $S_i$,
$Q_{i0}= \sqrt{ n_i S_i \over S_1} Q_{10}$, $E_{i0}=
{Q_{i0} \over S_i}=\sqrt{ n_i S_1 \over S_i}E_{10}$. 
%\cite{ref4},
%\begin{equation}Q_{i0}= \sqrt{ n_i S_i \over S_1} Q_{10}\qquad , \qquad E_{i0}=
%{Q_{i0} \over S_i}=\sqrt{ n_i S_1 \over S_i}E_{10}\ \ .
%\label{ec4}\end{equation}

We also take into account 
the evolution of the field and the charge
with the decay of the cluster,
$E_i=E_{i0}{ 1 \over (1+ {\tau \over \tau_{i0}})^2}$, 
$Q_i= Q_{i0} { 1 \over (1+ {\tau \over \tau_{i0}})^2}$
where $\tau \sim 1/ \sqrt{E_{i0}}$.
%Taking into account the evolution of the field and the charge
%with the decay of the cluster,
%
%\begin{equation} E_i=E_{i0}{ 1 \over (1+ {\tau \over \tau_{i0}})^2} \qquad ,
%\qquad
%Q_i= Q_{i0} { 1 \over (1+ {\tau \over \tau_{i0}})^2}\ \ ,
%\end{equation}
%where $\tau \sim 1/ \sqrt{E_{i0}}$,
%equation (\ref{ec1}) transforms into 
%\begin{equation}{dN_{q- \bar q} \over dy} \propto
%\sum_{i=1}^M Q_{i0}\int_0^\infty dx x
%{ 1 \over (1+ x)^4}\sum_{n=1}^\infty
%{1 \over n^2} \exp \Big [  -{\pi n m_j^2\over |g_j E_{i0}|}
%(1+ x)^2\Big ]\ \ ,
%\label{ec3}\end{equation}
%where $M$ is the total number of clusters.
\begin{figure}[t]
\centering\leavevmode
\epsfxsize=3.5in\epsfysize=1.75in\epsffile{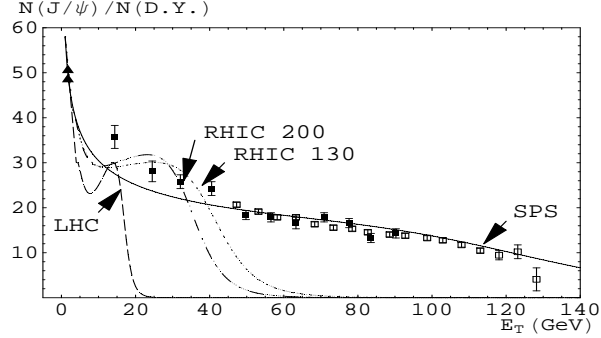}
\caption{Ratio of $J/\psi$ to DY events vs. $E_T$
%as predicted by (\ref{ec10}) and
%(\ref{ec11})
%for Pb-Pb collisions 
at SPS (solid line),
%Au-Au collisions at 
RHIC (dotted and dashed-dotted line)
%(130 GeV/n) (dotted line),
%Au-Au collisions at RHIC energies (200 GeV/n) (dashed-dotted line)
and 
%Pb-Pb collisions at 
LHC energies (dashed line).
%as a function of $E_T$. 
Arrows mark the percolation critical points.}
%and
%experimental data for SPS are from NA50 Collaboration [8,26].}
\label{figure1}
\end{figure}

\subsection{Probability of QGP formation}
%We include the probability of QGP plasma formation in a very simple
%continuum two-dimensional percolation model. 
%In an event with $\nu$ elementary
%collisions,
%there is a probability $P_{np}(\nu )$ of percolation not to occur, so a
%probability $1-P_{np}(\nu )$ of percolation to occur. 
It is assumed that the
$J/\psi$ is formed only in events in which there is not percolation.
The probability of percolation is taken to be 
%\cite{ref16}
\begin{equation}P_{perc}= 1/(1+\exp (-(\eta -\eta_c)/a)) \end{equation}
\noindent where $\eta = \pi r_0^2 N_s /A(\nu )$ corresponds to the
 density of strings,
$\eta_c=1.15$ is the critical density for percolation and $a=0.04$.
The probability for DY and $J/\psi$ production 
%to find a Drell-Yan event 
when there are $\nu$ elementary
collisions are
\begin{equation} P(DY|\nu )= \alpha_{DY} \;\nu\ \ , \end{equation}
%while for $J/\psi$ production one has
\begin{equation} P(J/\psi |\nu )= \alpha_{J/\psi}\;\nu\; e^{-\sigma \eta /  \pi
r_0^
2}\;
{1 \over \exp ({\eta-\eta_c \over a})+1}\ \ , \end{equation}
and the ratio $R$ of $J/\psi$ to DY events is
\begin{equation}R=k \exp (-\sigma \eta /
\pi r_0^2)/[\exp ({(\eta-\eta_c) /a})+1]\ \ .
\label{ec10}\end{equation}
However, as we said before, the probability of $c \bar c$ production increases
with
$\eta$. In order to include this fact, we multiply the expression (\ref{ec10})
by
\begin{equation}P_{c\bar c}(\eta )/P_{c\bar c}(\eta_{SPS}
)\label{ec11}\end{equation}
assuming that the leading $c \bar c$ pair production 
%for central Pb-Pb
%collisions at
% SPS
energies is given by the
Schwinger mechanism (\ref{ec1}).
%, expression (\ref{ec3}) and (\ref{ec4}).
%
%In figure \ref{figure1} 
Our results %for the ratio $J/\psi/DY$
%as a function of $E_T$ 
are plotted in Fig. \ref{figure1}. 
%It is seen that at RHIC and LHC energies an increase appears just
%before 
%the percolation critical point which is marked by arrows.
% in fig.
%\ref{figure1}.
%\begin{figure}[t]
%\centering\leavevmode
%\epsfxsize=3in\epsfysize=1in\epsffile{figure1.eps}
%\caption{Ratio of $J/\psi$ to Drell-Yan events vs. $E_T$
%%as predicted by (\ref{ec10}) and
%%(\ref{ec11})
%for Pb-Pb collisions at SPS energies (solid line),
%Au-Au collisions at RHIC energies (130 GeV/n) (dotted line),
%Au-Au collisions at RHIC energies (200 GeV/n) (dashed-dotted line)
%and Pb-Pb collisions at LHC energies (dashed line).}
%%as a function of $E_T$. Arrows mark the percolation critical points and
%%experimental data for SPS are from NA50 Collaboration [8,26].}
%\label{figure1}
%\end{figure}

%\section*{Acknowledgments}
%This is where one places acknowledgments for funding bodies etc.
%Note that there are no section numbers for the Acknowledgments, Appendix
%or References. Furthermore, the system will automatically generate the heading for 
%the reference section.

\end{document}